\documentclass{amsart}

\usepackage{epsfig}
\usepackage{latexsym}
\usepackage{amsmath}
\usepackage{mathrsfs}
\newtheorem{theorem}{Theorem}[section]
\newtheorem{proposition}[theorem]{Proposition}

\newcommand{\T}{\mathcal T}

\newcommand{\PP}{{\mathbb P}}
\newcommand{\EE}{{\mathbb E}}

\newcommand{\A}{{\mathscr A}}
\newcommand{\B}{{\mathscr B}}
\newcommand{\C}{{\mathscr C}}
   
\newcommand{\old}[1]{{}}

\title[Markovian log-supermodularity in phylogenetics]{Markovian log-supermodularity, and its applications in phylogenetics}
\author{Mike Steel and Beata Faller}

\thanks{We thank the New Zealand Marsden Fund for supporting this work}

\address{Allan Wilson Centre for Molecular Ecology and Evolution, Department of Mathematics and
  Statistics, University of Canterbury, Christchurch, New Zealand}

\email{m.steel@math.canterbury.ac.nz}



\begin{document}
 \begin{abstract}
We establish a log-supermodularity property for probability
distributions on binary patterns observed at the tips of a tree that
are generated under any $2$--state Markov process. We illustrate the
applicability of this result in phylogenetics by deriving an
inequality relevant to estimating expected future phylogenetic
diversity under a model of species extinction. In a further
application of the log-supermodularity property, we derive a purely
combinatorial inequality for the parsimony score of a binary
character. The proofs of our results exploit two classical theorems
in the combinatorics of finite sets.
\end{abstract}
\maketitle

\bigskip
\section{Introduction}

Finite-state Markov processes on trees are widely used in
evolutionary biology to model the way in which discrete
characteristics of present-day species have evolved from the state
present in some common ancestor \cite{fel, sem}.  In this paper, we
investigate a generic inequality that applies to $2$--state Markov
processes on trees, and provide two applications.

The first application, which was the motivation for our study, is to
the theory of biodiversity conservation. We consider the expected
loss of `phylogenetic diversity' under a model in which extinction
risk is associated with an underlying state that evolves on the
tree. We are interested in comparing this expected loss to simpler
models in which extinction events are treated independently; we find
that that when extinction events reflect phylogenetic history, then
the expected loss of phylogenetic diversity is always greater than
or equal to that predicted by an independent extinction scenario.
Essentially, this is because the probability that an entire `clade'
(the set of present-day species descended from a vertex in the tree)
becomes extinct is higher when the evolution of the influencing
state is taken into consideration than when we treat extinctions as
independent events.

In a second application, we derive a new, purely combinatorial
result concerning the `parsimony score' of a binary character on a
tree. We also briefly discuss how the generic inequality for
$2$--state Markov processes relates to recent work on phylogenetic
invariants and inequalities for particular submodels.

\section{Markov processes on trees}
Let $T = (V_T, E_T)$ be a tree with leaf set $X$.  Consider a Markov
random field on $T$ with state space $\{0,1\}$, and for each vertex
$v$ of $T$, let $\xi(v)$ be the random state ($0$ or $1$) that $v$
is assigned. This process is usually described as follows. We have a
root vertex $\rho$ for which we specify a probability, say $\pi_i$,
that $\xi(\rho)=i$, for $i \in \{0,1\}$. Direct all the edges of $T$
away from $\rho$ and for any arc $(r,s)$ of the resulting directed
tree $T=(V_T, A_T)$, let $P^{(r,s)}$ denote the $2 \times 2$
transition matrix for which the $ij$--entry (for $i,j \in \{0,1\}$)
is the conditional probability that $\xi(s) =j$ given that
$\xi(r)=i$. Specifying $\pi = [\pi_0, \pi_1]$ together with the
transition matrices $P^{(r,s)}$ for all the arcs $(r,s)$ of $\T$
uniquely defines the Markov random field on $T$ (see, for example,
\cite{cha, sem, ste}); an explicit formula appears below
(Eqn.~\ref{product}). We will assume throughout that $\pi$ is
strictly positive and that the following condition holds on each of
the transition matrices:
$$\det P^{(r,s)} \geq 0.$$
Notice that this determinant condition automatically holds if one
views the transition matrix for an arc as describing the net effect
of a continuous-time Markov process operating for some duration for
that arc. Note however that we are not assuming that any such process is
the same between the arcs of $T$ (i.e. the model is not necessarily stationary).

For $U \subseteq V_T$ let $P(U)$ denote the probability that $U$ is
precisely the set of vertices of $T$ in state $0$; that is:
$$P(U)= \PP(\{v \in V_T: \xi(v) = 0\} = U).$$
To express $P(U)$ in terms of the transition matrices and $\pi$,
let:
\begin{align*}
\delta(U,v) = \begin{cases} 0, & \mbox{if $v \in U$;}
\\ 1, &
\mbox{if $v \in V_T-U$.} \end{cases} \end{align*} Then, the Markov property gives:
\begin{equation}
\label{product}
P(U) = \pi_{\delta(U, \rho)}\cdot \prod_{(r,s) \in A_T}P^{(r,s)}_{\delta(U,r)\delta(U,s)}.
\end{equation}
For any subset $W$ of $X$ (the leaf set of $T$), let $p_W$ denote
the probability that $W$ is precisely the set of leaves of $T$ that
are in state $0$.  This marginal probability is given by:
\begin{equation}
\label{marginal}
p_W = \sum_{U \in \A_W}P(U)
\end{equation}
where:
\begin{equation}
\label{Asets}
\A_W := \{U \subseteq V_T:
U \cap X = W\}.
\end{equation}

An example to illustrate this concept is provided in Fig. 1.
\begin{figure}[ht] \begin{center}
\resizebox{5.5cm}{!}{
\input{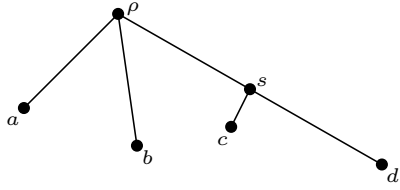}
}\caption{In this example, if $W=\{a,c\}$, then $p_W = \pi_0P_{00}^{(\rho,a)}P_{01}^{(\rho,b)}P_{00}^{(\rho,s)}P_{00}^{(s,c)}P_{01}^{(s,d)}+
\pi_0P_{00}^{(\rho,a)}P_{01}^{(\rho,b)}P_{01}^{(\rho,s)}P_{10}^{(s,c)}P_{11}^{(s,d)}+
\pi_1P_{10}^{(\rho,a)}P_{11}^{(\rho,b)}P_{10}^{(\rho,s)}P_{00}^{(s,c)}P_{01}^{(s,d)}+
\pi_1P_{10}^{(\rho,a)}P_{11}^{(\rho,b)}P_{11}^{(\rho,s)}P_{10}^{(s,c)}P_{11}^{(s,d)}$}
\end{center}
\label{figure1}
\end{figure}

A number of authors have noticed that certain inequalities hold for
quadratic functions of the $p_W$ values. For example, for any $x,y
\in X$ with $x\neq y$, it is well known that:
$$p_{\{x\}}\cdot p_{\{y\}} \leq  p_{\{x,y\}} \cdot p_\emptyset.$$
Moreover, in \cite{pea} the following inequality was described: for
subsets $\{x,y\}$ and $\{x,z\}$ of $X$ where $x,y,z$ are distinct,
we have
$$p_{\{x,y\}} p_{\{x,z\}} \leq p_{\{x,y,z\}} p_{\{x\}}.$$
The following proposition shows that these are special cases of a
much more general inequality.

\begin{proposition}
\label{subadd}
 For any $2$--state Markov process on a tree with leaf
set $X$, and any two subsets $Y,Z$ of $X$, we have:
$$p_Y\cdot p_Z \leq p_{Y \cup Z}\cdot p_{Y\cap Z}.$$
\end{proposition}
\begin{proof}
Let $A, B$ be arbitrary subsets of $V_T$. We first establish the following:
\begin{equation}
\label{ineq1}
P(A)\cdot P(B) \leq P(A \cup B) \cdot P(A \cap B).
\end{equation}
Applying Eqn. (\ref{product}) to $U \in \{A, B, A\cup B, A\cap B\}$,
the product $P(A)\cdot P(B)$ and the product $P(A \cup B) \cdot
P(A\cap B)$ can each be written as a product of two entries of $\pi$
multiplied by a product over the arcs $(r,s)$ of $T$ of two entries
of $P^{(r,s)}$. Moreover, regardless of where $r$ and $s$ lie in
relation to the sets $A,B$, the product of the two $\pi$ terms agree
in $P(A)\cdot P(B)$ and $P(A \cup B) \cdot P(A\cap B)$ (i.e.,  we
have $\pi_{\delta(A,\rho)}\pi_{\delta(B,\rho)} = \pi_{\delta(A\cup
B,\rho)}\pi_{\delta(A \cap B,\rho)})$, while the product of the two
$P^{(r,s)}$ terms agree in $P(A)\cdot P(B)$ and $P(A \cup B) \cdot
P(A\cap B)$, except for the cases in which either (i) $r \in A-B$
and $s \in B-A$, or  (ii) $r \in B-A$ and $s \in A-B$. However, in
both case (i) and (ii), the product $P^{(r,s)}_{01}P^{(r,s)}_{10}$
appears in the term for $P(A)\cdot P(B)$ while
$P^{(r,s)}_{00}P^{(r,s)}_{11}$ appears in the term for $P(A \cup B)
\cdot P(A\cap B)$, and the former term is less or equal to the
second since
$$P^{(r,s)}_{00}P^{(r,s)}_{11}- P^{(r,s)}_{01}P^{(r,s)}_{10} = \det P^{(r,s)}$$
and $\det P^{(r,s)} \geq 0$ by assumption.  Consequently, all the
terms in $P(A)\cdot P(B)$ are either less than or equal to (in cases
(i) and (ii)), or equal to (in all remaining cases) the
corresponding terms in $P(A \cup B) \cdot P(A\cap B)$. This
establishes (\ref{ineq1}).

We now invoke a classical result of Ahlsewede and Daykin (1978)
\cite{ahl}, sometimes called the `four functions theorem'. A
particular form of this theorem that suffices for our purposes is
the following (we follow \cite{and}).  Suppose we have a finite set
$S$ and a function $\alpha$ that assigns a non-negative real number
to each subset of $S$. Suppose that $\alpha$ satisfies the property
that for all subsets $A,B$ of $S$
$$\alpha(A)\alpha(B) \leq \alpha(A \cup B)\alpha(A \cap B).$$
For a collection $\C$ of subsets of $S$, let $\alpha(\C) := \sum_{C
\in \C}\alpha(C).$ Then for any two  collection of subsets  of $S$,
$\A$ and $\B$, say, we have:
\begin{equation}
\label{4fun}
\alpha(\A)\alpha(\B) \leq \alpha (\A \vee \B)\alpha(\A \wedge \B),
\end{equation}
where $$\A \vee \B := \{E \subseteq S: E = A \cup B: A \in \A, B \in \B\}, \mbox{ and }$$
$$\A \wedge \B := \{E \subseteq S: E = A \cap B: A \in \A, B \in \B\}.$$
We will apply this to our problem by taking $S= V_T, \alpha(U) = P(U)$ and noting that
$\alpha$ satisfies the required hypothesis by (\ref{ineq1}).  Recall the definition of
$\A_W$ in (\ref{Asets}) and note that:
$$\A_Y \vee \A_Z = \A_{Y \cup Z}, \mbox{ and } \A_Y \wedge \A_Z = \A_{Y \cap Z}.$$
Thus taking $\A=\A_Y$ and $\B = \A_Z$ in (\ref{4fun}) we deduce
that:
$$\alpha(\A_Y)\alpha(\A_Z) \leq \alpha(\A_{Y \cup Z})\alpha(\A_{Y \cap Z}).$$
The proposition now follows by observing that $p_W = \alpha(\A_W)$ for all subsets $W$ of $X$, in particular
the subsets $Y, Z, Y \cup Z$ and $ Y \cap Z$.

\end{proof}

\section{Applications in phylogenetics}

We now describe somee applications of Proposition~\ref{subadd}.

\subsection{Expected future phylogenetic diversity}

We first show how Proposition~\ref{subadd}, together with another inequality, provides a general inequality concerning the loss of expected future biodiversity under species extinction models.

Suppose that $T$ is a rooted tree with leaf set $X$, and with each
arc $e=(u,v)$ of $T$ there is an associated length $\lambda_e$.
Given a subset $Y$ of $X$, the {\em phylogenetic diversity} (PD) of
$Y$, denoted $\varphi_Y$, is the sum of the lengths of the edges of
the minimal subtree of $T$ connecting the root and the leaves in
$Y$. Under various possible interpretations of the $\lambda$ values,
PD has been widely used as a measure for quantifying present and
expected future biodiversity \cite{fai, fal, nee}.

For each species $x \in X$ let $E_x$ denote the event that species
$x$ is extinct at some future time $t$.  Then the expected
phylogenetic diversity of the set of species that are extant at time
$t$, referred to as {\em expected future PD} and denoted
$\EE[\varphi]$, is given by:
\begin{equation}
\label{var}
\EE[\varphi]= \sum_{e=(u,v) \in A_T} \lambda_e\cdot(1-\PP(\bigcap_{x \in C_v}E_x))= \varphi_X-\sum_{e=(u,v) \in A_T} \lambda_e\cdot\PP(\bigcap_{x \in C_v}E_x),
\end{equation}
where $C_v$ denotes the subset of $X$ that is separated from the
root by $v$. A simple model, referred to as the {\em generalized
field of bullets model} (g-FOB) in \cite{fal} (generalizing an
earlier model from \cite{nee}), assumes that the events $E_x$ are
independent. Then, if we let $p_x=\PP(E_x)$, the value of
$\PP(\bigcap_{x \in C_v}E_x)$ in (\ref{var}) (the probability of the
extinction of all the species descended from $v$) is given by:
\begin{equation}
\label{peqs}
\PP(\bigcap_{x \in C_v}E_x) =  \prod_{x \in C_v}p_x.
\end{equation}

An example to illustrate this concept is provided in Fig. 2.

\begin{figure}[ht] \begin{center}
\resizebox{6.6cm}{!}{
\input{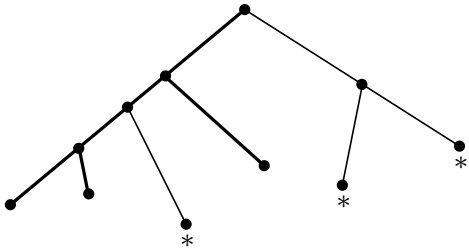}
} \caption{If the species indicated by * become extinct, then the remaining phylogenetic diversity is the sum of the lengths of the bold edges indicated.}
\end{center}
\label{figure2}
\end{figure}

The assumption that the events $E_x$ are independent is likely to be
unrealistic in most settings (see, for example, \cite{hea, sim}).
For example, species that are `close together' in $T$ are more
likely to share attributes that may put them at risk in a hostile
future environment.

To take a simple but topical scenario, consider extinction risk due
to climate change. Suppose that the extinction risk of each species
in $X$ is partially influenced by some associated binary state ($0$
or $1$) where state $0$ confers an elevated risk of extinction under
climate change.  We suppose that these states are not known in
advance for the species in $X$, and that this state has evolved
under some Markovian model on $t$. Once the states are determined at
the leaves, then extinction proceeds according to the g-FOB model,
where species $x$ is extinct at time $t$ with probability $p_x^i$ if
it is in state $i \in \{0,1\}$.  We call this a {\em state-based
field of bullets} model (s-FOB). Note that this includes the g-FOB
model as a special case where $p_x^0=p_x^1$ for all $x$. Moreover,
once we condition on the state for each leaf, an s-FOB model is just
a g-FOB model with modified extinction probabilities, but we are
assuming that these states are unknown (in line with the uncertainty
over what features may be helpful for an organism in a future
climate).

With any s-FOB model we also have an associated g-FOB model in which
the extinction probability of each species $x$ is the same as in the
s-FOB model. That is, in the g-FOB model we set:
\begin{equation}
\label{px}
p_x = p_x^0\PP(\xi(x)=0) + p_x^1\PP(\xi(x)=1),
\end{equation}
 where $\xi$ describes the
Markov process for the binary character. A natural question arises:
how does the future expected PD of an s-FOB model compare with that
of its associated g-FOB model? The following result provides a
general inequality.

\begin{theorem}
Consider a fixed tree with branch lengths and leaf set $X$. Consider
an s-FOB model, in which state $1$ is advantageous for each species,
i.e., $p_x^1 \leq p_x^0$ for all $x \in X$. Then the expected future
PD of this model is less or equal to the expected future PD of the
associated g-FOB model.
\end{theorem}
\begin{proof}

In view of (\ref{var}) and (\ref{peqs}), it suffices to show that:
\begin{equation}
\label{ine1}
\prod_{x \in C_v} p_x \leq \PP(\bigcap_{x \in C_v}E_x) ,
\end{equation}
where
$p_x$ is defined by Eqn. (\ref{px}).

For each subset $W$ of $C_v$ let $p_W$ denote the probability that
the set of elements of $C_v$ in state $0$ is precisely $W$.  Then:
$$\PP(\bigcap_{x \in C_v}E_x) = \sum_{W \subseteq C_v} p_W\prod_{x
\in W}p_x^0 \prod_{x \in C_v - W} p_x^1.$$ Thus, if we let:
\begin{align*}
f_x(W) = \begin{cases} p_x^0, & \mbox{if $x \in W$;}
\\ p_x^1, &
\mbox{if $x \in C_v-W$.} \end{cases} \end{align*} then:
$$\PP(\bigcap_{x \in C_v}E_x) = \sum_{W \subseteq C_v} p_W\prod_{x \in C_v}f_x(W).$$
Moreover:
$$p_x = p_x^0\PP(\xi(x)=0) + p_x^1\PP(\xi(x)=1) = \sum_{W \subseteq C_v} p_W f_x(W),$$ where the second
equality arises by considering in the summation those $W$ containing
$x$ and those not containing $x$.
 Consequently, (\ref{ine1}) is equivalent to the
requirement that:
\begin{equation}
\label{ine2}
\prod_{x \in C_v} \left( \sum_{W \subseteq C_v} p_W f_x(W)\right) \leq \sum_{W \subseteq C_v} p_W\prod_{x \in C_v}f_x(W).
\end{equation}
The proof of (\ref{ine2}) involves combining
Proposition~\ref{subadd} with the  FKG inequality of Fortuin,
Kasteleyn and Ginibre (1971) \cite{fkg}, a particular (and
multivariate) form of which we now recall.

Given a finite set $S$, suppose that $f_1, f_2, \ldots, f_n$ are
functions from the power set of $S$ into the non-negative real
numbers, and that these satisfy the condition:
\begin{equation}
\label{inclusion}
A \subseteq B \Rightarrow f_i(A) \leq f_i(B).
\end{equation}
Furthermore, suppose that $\mu$ is a probability measure on the
subsets of $S$ which satisfies the log-supermodularity condition:
\begin{equation}
\label{inclusion2}
\mu(A)\mu(B) \leq \mu(A \cup B)\mu(A \cap B).
\end{equation}
Then:
\begin{equation}
\label{ine3}
\prod_{i=1}^n\left(\sum_A\mu(A)f_i(A)\right) \leq
\sum_A \mu(A)\prod_{i=1}^n f_i(A)
\end{equation}
 where the
summations are over all subsets of $S$.

We apply this form of the FGK inequality  by
taking $S = \{1,\ldots, n\} = C_v$, $\mu(W) = p_W$, and $f_x$ as defined above.
Then $f_x$ satisfies (\ref{inclusion}) by the hypothesis that $p_x^1 \leq p_x^0$ for all $x$,
while $\mu$ satisfies (\ref{inclusion2}) by Proposition 2.1.
Then inequality (\ref{ine3}) provides the required inequality (\ref{ine2}).
This completes the proof.

\end{proof}

\subsection{Combinatorics of parsimony}
We now describe a further application of Proposition~\ref{subadd} by
deriving a purely combinatorial result concerning a measure
(`parsimony score') that underlies certain approaches for inferring
evolutionary history (see, for example, \cite{fel}).

Given a function  $f:X \rightarrow \{0,1\}$, recall that the {\em
parsimony score} of $f$ on $T$, denoted $l(f,T)$, is the minimum
number of edges that have different states assigned to their
endpoints, across all extensions $F:V_T \rightarrow \{0,1\}$ of $f$.
For example, for the tree $T$ in Fig. 1, and the function $f$
defined by $f(a)=f(c)=0, f(b)=f(d)=1$, we have $l(f, T) =2$; for
this example, there are two minimal extensions $F$ of $f$
corresponding to $F(\rho)=F(s)=i$ for $i \in \{0,1\}$ (for further
details concerning the mathematical properties of parsimony score,
see \cite{sem}).
 For $W \subseteq X$, let $f_W$ denote
the function that assigns state $0$ to the elements of $W$, and
assigns state $1$ to the elements of $X-W$. The following result
states that the parsimony score function for a given tree is
submodular.
\begin{theorem}
\label{parsy}
 For any tree $T$ with leaf set $X$ and subsets $Y, Z$,
of $X$ we have:
$$
l(f_Y, T)+l(f_Z,T) \geq l(f_{Y\cup Z}, T)+l(f_{Y\cap Z}, T).
$$
\end{theorem}
\begin{proof}
Consider the $2$--state Markov random field on $T$ with
$\pi_0=\pi_1=0.5$, and set each transition matrix $P^{(r,s)}$ to be
the symmetric $2 \times 2$ matrix with off-diagonal entry
$\epsilon>0$. Then, for any $W \subseteq X$ a straightforward
calculation shows that:
\begin{equation}
\label{limit}
p_W = C_W\epsilon^{l(f_W, T)}(1 + o(\epsilon)),
\end{equation}
for a constant $C_W$ that depends only on $W$ and $T$ and not
$\epsilon$ (specifically, $C_W$ is the number of minimal extensions
of $f_W$ to the vertices of $T$ multiplied by $\frac{1}{2}$). Now
Proposition~\ref{subadd}, expressed using logarithms, states that:
\begin{equation}
\label{ineq}
-\log(p_Y)-\log(p_Z) \geq -\log(p_{Y
\cup Z}) -\log(p_{Y \cap Z}).
\end{equation}
Applying (\ref{limit}) (and noting that $\log(1+o(\epsilon)) =
o(\epsilon)$), the left-hand side of (\ref{ineq}) is:
$$(l(f_Y,T)+ l(f_Z,T))\log\left(\frac{1}{\epsilon}\right) -
\log(C_YC_Z) + o(\epsilon)$$ while the right-hand side of
(\ref{ineq}) is: $$(l(f_{Y \cup Z},T) + l(f_{Y \cap
Z},T))\log\left(\frac{1}{\epsilon}\right) - \log(C_{Y\cup Z}C_{Y
\cap Z}) + o(\epsilon).$$ Theorem~\ref{parsy} now follows by letting
$\epsilon$ tend to zero.
\end{proof}

\section {Concluding Remarks}
\begin{itemize}
\item[(i)]
It is possible to establish  Theorem~\ref{parsy} using a purely combinatorial proof, by
first invoking Menger's theorem from graph theory to handle the case where $Y$ and $Z$ are disjoint, and then using a complementation argument for the case $Y\cup Z = X$. The remaining case where $X-(Y \cup Z)$ is non-empty can then be established by a somewhat detailed argument that uses induction on $|X|$.

\item[(ii)]
Proposition~\ref{subadd} provides a collection of polynomial
inequalities on the $p_W$ values, which have recently been
studied for a particular class of Markov $2$--state models in
\cite{mat}. These polynomial inequalities complement the
much-studied `phylogenetic invariants' (polynomial identities in
the $p_W$ values), which hold under various restrictions on the
Markov model.  Combining these phylogenetic invariants with the
polynomial inequalities provides a way of characterizing when a
probability distribution arises on a tree under a Markov process
(either with or without restrictions). For the 2-state Markov
process on a tree with $3$ leaves, this was solved in
\cite{pea}.

\item[(iii)] It may be of interest to derive an extension of Proposition~\ref{subadd} that
applies when the state space has a size greater than $2$.
\end{itemize}


\begin{thebibliography}{99}

\bibitem{ahl}
R. Ahlswede,  and D.E. Daykin,  An inequality for the weights of two
families of sets, their unions and intersections, {\em Z. Wahrsch.
V. Geb.} {\bf 43} 183-185 (1978).

\bibitem{and} I. Anderson,  {\em Combinatorics of finite sets}, Dover
Publications, New York, (1987).


\bibitem{cha} J.T. Chang,  Full reconstruction of
  Markov models on evolutionary trees: identifiability and
  consistency, {\em Math, Biosci.} {\bf 137} 51--73 (1996).

\bibitem{fai} D.P.  Faith,  Conservation evaluation and phylogenetic
diversity, {\em Biol. Conserv.} {\bf 61} 1–-10 (1992).


\bibitem{fal}
B. Faller, F. Pardi and M. Steel,  Distribution of phylogenetic
diversity under random extinction, {\em J. Theor. Biol.} {\bf 251}
286--296 (2008).

\bibitem{fel}
J. Felsenstein,{\em Inferring phylogenies}, Sinauer Press, (2003).

\bibitem{fkg}
C.M. Fortuin, P.W. Kasteleyn and J. Ginibre, Correlation
inequalities on some partially ordered sets, {\em Commun. math.
Phys.} {\bf 22} 89-103 (1971).

\bibitem{hea} S.B. Heard, and A.O. Mooers,  Phylogenetically
patterned speciation rates and extinction risks change the loss of
evolutionary history during extinctions, {\em Proc. R. Soc. Lond.
B.} {\bf 267} 613--620 (2000).

\bibitem{mat} F.A. Matsen, Fourier transform inequalities for phylogenetic trees.
arXiv:0711.3492 (q-bio.PE) (2008).


\bibitem{nee} S. Nee and R.M. May,  Extinction and the
loss of evolutionary history, {\em Science} {\bf 278}(5338) 692–-694
(1997).


\bibitem{pea} J. Pearl and M. Tarsi Structuring causal trees,
{\em J.  Complexity}, {\bf 2}, 60--77 (1986).


\bibitem{sem} C. Semple and M. Steel, {\em Phylogenetics}, Oxford
University Press, Oxford, (2003).

\bibitem{sim} H. Simianer,  Accounting for non-independence of extinction probabilities in the derivation of
 conservation priorities based on Weitzman's diversity concept, {\em Conserv. Genet.} {\bf 9},
 171--179 (2008).

\bibitem{ste} M. Steel  Recovering a tree from the leaf
  colourations it generates under Markov model. {\em Appl. Math. Lett.}, {\bf 7} 19--23 (1994).


\end{thebibliography}
\end{document}